# Structure and dynamics of core/periphery networks

PETER CSERMELY*

*Department of Medical Chemistry, Semmelweis University, P.O. Box
260. H-1444 Budapest, Hungary*
*Corresponding author: Peter.Csermely@med.semmelweis-univ.hu

ANDRÁS LONDON

*Department of Computational Optimization, University of Szeged, P. O. Box 652. H- 6701, Szeged, Hungary*

LING-YUN WU

*Institute of Applied Mathematics, Academy of Mathematics and Systems Science, Chinese Academy of Sciences, No. 55, Zhongguancun East Road, Beijing 100190, China*

AND

BRIAN UZZI

*Northwestern University Institute on Complex System (NICO), Northwestern University, Evanston, Illinois 60208, USA; Kellogg School of Management, Northwestern University, Evanston, Illinois 60208, USA*

Recent studies uncovered important core/periphery network structures characterizing complex sets of cooperative and competitive interactions between network nodes, be they proteins, cells, species or humans. Better characterization of the structure, dynamics and function of core/periphery networks is a key step of our understanding cellular functions, species adaptation, social and market changes. Here we summarize the current knowledge of the structure and dynamics of "traditional" core/periphery networks, rich-clubs, nested, bow-tie and onion networks. Comparing core/periphery structures with network modules, we discriminate between global and local cores. The core/periphery network organization lies in the middle of several extreme properties, such as random/condensed structures, clique/star configurations, network symmetry/asymmetry, network assortativity/disassortativity, as well as network hierarchy/anti-hierarchy. These properties of high complexity together with the large degeneracy of core pathways ensuring cooperation and providing multiple options of network flow re-channelling greatly contribute to the high robustness of complex systems. Core processes enable a coordinated response to various stimuli, decrease noise, and evolve slowly. The integrative function of network cores is an important step in the development of a large variety of complex organisms and organizations. In addition to these important features and several decades of research interest, studies on core/periphery networks still have a number of unexplored areas.

*Keywords:* bow-tie networks, core/periphery networks, nested networks, onion networks, rich-club networks.

# 1. Introduction

Intuitively, the concept of the network core usually refers to a central AND densely connected set of network nodes, while the periphery of the network denotes a sparsely connected, usually non-central set of nodes, which are linked to the core. (The "AND" is important in the above intuitive definition, since all nodes of the core are rather central, but by far not every set of central nodes forms a network core.) The concept of the network core may be approached from many directions (including various core defining algorithms; rich-clubs referring to an interconnected set of network hubs; nested networks; the bow-tie structures of directed networks, as well as the highly robust onion network structures [1-10]), and therefore has many types of definitions, which we will detail and compare in Section 2 of our review.

Several observations of network dynamics implied the development and utilization of core/periphery network structures. The early work of Ramon Margalef in 1968 [11] emphasized the role of asymmetry and heterogeneity of complex systems. The seminal 1972 paper of Robert May [12] proposed that network stability may be achieved either by the development of a nested-like core/periphery structure, or by network modules. Later studies confirmed that network cores facilitate system robustness and evolvability helping the adaptation to large fluctuations of the environment, as well as to noise of intrinsic processes. The network core can be regarded as a highly degenerate segment of the complex system, where the densely intertwined pathways can substitute and/or support each other (Fig. 1; [7-9,13]). Engineering processes and engineered products usually have a core/periphery structure, such as that of manufacturing assembly processes or the core of low-level firmware (e.g. the hardware of the device) combined with the periphery of high-level firmware (e.g. the operational instructions or software of the device). Even money can be thought of as a network core of multiple economic processes [9,14]. A special type of core/periphery networks, onion networks emerged as the most robust structures against simultaneous random and targeted attacks [10,15,16]. Changes of system resources maintaining network connections and/or interaction constraints may lead to topological phase transitions of networks. Core/periphery structures are often formed as a response of complex systems to various types of stresses or crisis conditions [17-22]. We will describe the dynamics leading to the development of and characterizing core/periphery network structures in Section 3 of our review.

Importantly, based on the method of spectral scaling [23] Estrada proved analytically that every possible network can be only in one of four possible topological classes being either good expander (i.e. a sparsely populated but highly connected, homogeneous network with good communication properties and lacking bottlenecks), a network with modules, a core/periphery network or a network with holes [24].

Core/periphery structure was detected in many complex systems including protein structure networks; protein-protein interaction networks (interactomes), metabolic, signalling and gene regulatory networks; networks of immune and neuronal cells;

ecosystems; animal and human social networks and related networks, such as the World Wide Web or Wikipedia; engineered networks (such as the Internet, power-grids or transportation networks), as well as networks of the world economy. Flow-type networks (such as metabolic networks, signalling networks, the Internet, etc.) often develop a more characteristic core/periphery structure than association-type networks (such as protein-protein interaction networks, social networks, etc.) [2]. We will detail and compare the core/periphery structure of these networks in Section 4.

We conclude our introduction with a few general remarks on core/periphery networks.

- Null-models (i.e.: appropriately randomized networks giving a "default" value, which has to be compared with the value obtained from the real world network) have a key importance of the definition of network core properties [1,2,4-6,25]. Complex systems have many features (such as the probability of hubs), which are more extreme than expected by chance. This is also true for the emergence of network cores. However, the selection of an appropriate null-model is not an obvious task. Imposing too many constraints on the null-model decreases its power, and increases the chance of statistical errors (e.g. that the null-model will contain circular argumentation). Importantly, null-models require a correct interpretation of the generative processes of the randomized network assemblies. Null-models have to be tested regarding the related concepts of appropriate sampling, level of network homogeneity and occurrence of autocorrelation [25]. Additionally, the accurate comparison of null-model corrected core-periphery measures between networks is also a difficult task. We will detail the various null-models of core definitions in Section 2.
- The absolute and relative size of the core (i.e.: the number of nodes and edges forming the core and/or their ratio to the total number of nodes and edges in the network) is a key property of core/periphery structures. An extensive core was proposed to allow a larger flexibility and adaptability of the network [9]. However, the larger flexibility of a large core may come together with a presumed restriction of network controllability (in the sense of maximally achievable control) [9,26,27].
- Besides their size, the number of network cores may also vary. Cores of well separated network modules (i.e.: the set of their densest, or most belonging nodes and edges) may be regarded as multiple network cores [28-32]. Such a multi-core network has a cumulus structure resembling to the puffy, cumulus clouds on the sky. On the contrary, if network modules became less separated (fuzzier), the multiple network cores tend to disappear, and the network structure starts to resemble to that of stratus clouds. A stratus → cumulus network structure transition occurs, when the complex system experiences stress, crisis, or a decrease in resources [20]. Modular structures were also described for onion networks [33], where peripheral nodes are not only connected to core nodes but also to each other.
- We focus our review on core/periphery structures of network nodes. However, we would like to note that edge-cores may also be defined. Most central edges form a network skeleton, which is vital for the communication of the network. The network skeleton becomes an especially important and exciting concept in weighted and/or directed networks, which may show completely different behaviour than unweighted, and/or undirected networks [18,34,35]. However, most network edge-skeletons do not form a densely inter-connected network segment, and therefore do not conform to the intuitive concept of network cores. This is the reason, why we did not include edge-cores to the detailed discussion of our review.

**2. Definitions and structural properties of core/periphery networks**

We will start our review with the description of core/periphery network models, rich-clubs, nested network structures, the bow-tie organization of several directed networks and the robust onion network structures.

*2.1 Definitions of network cores and peripheries*

A number of local, dense network structures, such as cliques, $k$-clans, $k$-clubs, $k$-cliques, $k$-clique-communities, $k$-components, $k$-plexes, strong LS-sets, LS-sets, lambda sets, weak LS-sets or $k$-cores have been described from the late 1940s (see Table 1; [36-52]). The node-removal definition of $k$-cores proved to be especially powerful leading to the definition of several leaf-removing pruning algorithms [53-57] including sets of progressively embedded cores of directed networks [57]. However, many of these dense subgraphs characterize local network topology, were later used for the definition of network modules (or in other words: network communities) [28,29,42,45], and usually lacked the discrimination of the network periphery, i.e. the analysis of those nodes, which did not belong to the core. Peripheral nodes are usually not connected to each other, while nodes outside the dense subgraphs listed in Table 1 are often connected with each other. Additionally, networks usually have multiple modules, while they usually have only one core. Having said this we have to note that in the traditional use of the words there is no clear discrimination between network modules and network cores, since there are core/periphery type networks, called onion networks [10,15,16,33], where the peripheral nodes do connect each other, and multiple network cores were also described [31,33]. We will return to the definition of core/periphery networks in Section 2.6 and in our Conclusions at the end of the review.

The concept of network core and periphery emerged in different fields from the late 70s, like in social networks [41,58], in the context of scientific citation networks [59,60], or in networks related to the economy [61-63]. However, the core/periphery network structure was formally defined first only in the end of the 90s by Borgatti & Everett [1]. The discrete approach of Borgatti & Everett [1] was based on the comparison of the adjacency matrix of the network. In their concept the core is a dense network entity, which "can not be subdivided into exclusive cohesive groups or factions". Thus, an ideal core/periphery network model consist a fully-linked core and a periphery that is fully connected to the core, but there are no links between any two nodes in the periphery. Mathematically, let $G = (V,$ $E$) an undirected, unweighted graph with $n$ nodes and $m$ edges and let $A = (a_{ij})_{i,j}$ the adjacency matrix of $G$, where $a_{ij} = 1$ if node $i$ and node $j$ are linked, and 0, otherwise. Let $\delta$ be a vector of length $n$ with entries equal to one or zero, if the corresponding node belongs to the core or the periphery, respectively. Furthermore, let $\Delta = (\Delta_{ij})$ be the adjacency matrix of the ideal core/periphery network on $n$ nodes and $m$ edges, where $\Delta_{ij} = 1$ if $\delta_i = 1$ and $\delta_j = 1$, and 0 otherwise (i.e. $\Delta = \delta^T \delta$, where $\delta^T$ is the transpose of the row vector $\delta$). Determining how a network has a core-periphery structure is an optimization problem aiming to find the vector $\delta$ such that the expression

(1)
$$\rho = \sum_{i,j} A_{ij} \Delta_{ij}$$

achieves its maximum value. The measure $\rho$ is maximal, when the adjacency matrix $A$ and the matrix $\Delta$ are identical, hence a network has core/periphery structure, if $\rho$ is large [1]. The Borgatti-Everett algorithm finds the vector $\delta$ such that the correlation between the $\Delta = \delta^T \delta$ matrix and the data (adjacency) matrix $A$ is maximized. The continuous extension of this model defines $\Delta_{ij} \in [0, 1]$, if $\delta_i = 1$ or $\delta_j = 1$, and runs the same algorithm as in the discrete case [1].

Borgatti & Everett [1] already warned that "what is missing in this paper is a statistical test for the significance of the core/periphery structures found by the algorithm". This is an important note of the necessity of appropriate null-models what we emphasized in Section 1. Null-models are important all the more, since Chung & Lu [64] showed that power-law random graphs, in which the number of nodes of degree $k$ is proportional to $k^{-\beta}$, almost surely contain a dense subgraph what has short distance to almost all other nodes in the graph when the exponent $\beta \in [2, 3]$. Utilizing the power of null-models Holme defined a core/periphery coefficient in 2005 [2] using the extension of the closeness centrality [65] to a subset $U$ of the network nodes and including null-model graphs with the same degree sequence as the original one. The extended closeness centrality $C_C(U)$ for a subset of nodes $U (\subseteq V)$ is defined as

(2)
$$C_C(U) = \frac{1}{\mathrm{avg}_{i \in U}(d(i,j)_{j \in V \setminus \{i\}})}$$

where $d(i, j)$ is the distance between node $i$ and node $j$. Thus, the core/periphery coefficient is formally defined as

(3)
$$c_{cp}(G) = \frac{C_C(V_{k-core}(G))}{C_C(V(G))} - \mathrm{avg} \left[ \frac{C_C(V_{k-core}(G'))}{C_C(V(G'))} \right]_{G' \in \mathscr{G}(G)}$$

where $V(G)$ is the set of nodes of the original graph $G$, $V_{k\text{-}core}(G)$ is the set of nodes of the maximum subgraph of $G$ with minimum degree $k$ and maximal $C_C(U)$ value and finally, $\mathcal{G}(G)$ is an ensemble of graphs with the same degree sequence as $G$ [2].

Motivated by the continuous model of Borgatti and Everett [1], Rombach *et al.* [31] introduced a new method to investigate the core/periphery structure of weighted, undirected networks. Using the same notation as in equation (1) their idea was finding vector $\delta$'s components as a shuffle of a given vector $\delta^*$, whose components specify *local core* values by using a transition function to interpolate between core and periphery nodes

(4)
$$\delta_i^* = \frac{1}{1+\exp(-(m-n\beta)\tan(\pi\alpha/2))},$$

where the two parameters $\alpha, \beta \in [0, 1]$. $\alpha$ defined the sharpness of the boundary between the core and the periphery; the value zero being the fuzziest. Parameter $\beta$ sets the size of the core: as $\beta$ varied from 0 to 1, the number of nodes included in the core varied from $n$ to 0. As a second step the *core quality*, $R$ was defined as

(5)
$$R = \sum_{i,j} A_{ij} \delta_i \delta_j$$

and its maximum was found using simulated annealing. Finally, the total *core score* of node $i$ was defined as

(6)
$$CS(i) = Z \sum_{\alpha,\beta} \left( \delta_i^*(\alpha,\beta) \sum_{j \in N(i)} \delta_j^*(\alpha,\beta) \right)$$

where N($i$) were the neighbouring nodes of $i$, and $Z$ was a normalization factor chosen such that $\max_k|CS(k)| = 1$. Nodes were more likely to be part of a network's core both if they had high strength, and if they were connected to other nodes in the core. The latter idea was reminiscent of the ideas of eigenvector centrality and PageRank centrality, which recursively define nodes as important based on having connections to other nodes that are important. Their method could identify multiple cores in a network, i.e. parallel core/periphery and network community structures [31]. Cores of network modules have been identified by other methods too [28-30,32].

*2.2 Rich-clubs*

Rich-clubs (Fig. 2.) were first introduced by Zhou & Mondragón [3] finding that connection-rich nodes (i.e. hubs being in the top X% of the nodes with largest degree) of the Internet are inter-connected, and form a dense core of the network. They defined the *topological rich-club coefficient*, Φ(*k*), i.e. the proportion of edges connecting the rich nodes, with respect to all possible number of edges between them. Formally,

$$\phi(k) = \frac{E_{>k}}{\binom{N_{>k}}{2}} = \frac{2E_{>k}}{N_{>k}(N_{>k}-1)}$$

(7)

where $N_{>k}$ refers to the nodes having a degree higher than *k*, and $E_{>k}$ denotes the number of edges among the $N_{>k}$ nodes in the rich-club. In other words, Φ(*k*) measures the probability that two nodes with higher degree than *k* are actually linked. If Φ(*k*) = 0 the nodes do not share any links, if Φ(*k*) = 1 the rich-nodes forms a fully-connected subnetwork, a clique [3].

The initial concept of rich-clubs [3] seemed to be related to network assortativity, where similar-degree nodes are preferentially attached to each-other [66]. However, the core/periphery network structure is more related to disassortative rich-clubs, where the association of high-degree nodes is accompanied by the lack of similarly high number of edges between low-degree nodes [4]. Later, it was determined that the above, intuitive definition of the rich-club property holds predominantly for sparse networks. If the number of connections is sufficiently high, and the degree distribution is slowly decreasing, even random networks without multiple and self-connections contain a core of about $n^{2/3}$ highly interconnected nodes, where *n* is the number of nodes in the network [67]. This property of dense random networks and the difficulties of rich-club detection in other dense networks [68] warned for the use of appropriate null-models.

Colizza *et al.* [4] introduced the first null-model to detect rich-clubs using the randomization procedure of Maslov & Sneppen [69], which preserved the degree sequence of the original graph. Thus the rich-club coefficient was defined as

$$\rho(k) = \frac{\phi(k)}{\phi_{rand}(k)}$$

(8)

where $\Phi_{rand}(k)$ is the rich-club coefficient of the appropriately randomized benchmark graph. Colizza *et al.* [4] observed that Φ(*k*) monotonically increases in nodes with increasing degree even in the case of Erdős-Rényi random graphs [70], which confirmed that assessment of the rich-club property requires an appropriate null-model – especially in dense networks.

Soon after, Mondragón & Zhou [71] suggested the discrimination between the *rich-club coefficient* (Φ(*k*) as defined above), the *rich-club structure* (which is the rich-club coefficient measured across various levels of *k*), the rich-club phenomenon (which refers to the dynamic evolution leading to the development of rich-clubs) and the *rich-club ordering*, which relates the rich-club coefficient to an appropriate null-model. They argued that the rich-club structure may give important information on a network even without a null-model.

Additionally, the null-model can not identify the evolutionary process leading to a rich-club structure. They also introduced two other rewiring processes to create null-models. The first method preserved the rich-club coefficient as a function of the rank of the node in the original network, and resulted in network ensembles having a similar degree distribution and assortativity like those of the original network. This method defined a random network having the same number of nodes and edges as the original network. In each step a randomly selected edge was rewired. Then the square deviation, *d*, of the original rich-club coefficient and the rich-club coefficient of the randomized network

(9)
$$d = \sum_{k=1}^{N}(\phi(k) - \phi_{rand}(k))^2$$

was calculated, and the algorithm accepted the rewiring, if it reduced *d*. The second method preserved both the rich-club coefficient and the degree-distribution. This method selected a randomly chosen pair of edges. If these edges were assortatively wired concerning their degree, they were discarded, and a new pair was considered; otherwise the four end-nodes of the edges were reshuffled at random. This procedure was repeated for a large number of times [71].

Later, several extensions of the above definitions were given for weighted networks using weight-reshuffle and/or weight and edge-reshuffle null-models. Weighted networks may reveal a completely different rich-club structure than their unweighted pairs: formation of local dense groups in the absence of a global rich-club, as well as lack of cohesion in the presence of rich-club ordering [68,72,73]. Opshal *et al.* [72] proposed the weighted rich-club coefficient

(10)
$$\phi^w(k) = \frac{W_{>k}}{\sum_{l=1}^{E_{>k}} w_l^{rank}}$$

where the numerator is the total weight of edges connecting the $N_{>k}$ nodes, the denominator is the sum of weights of the $E_{>k}$ strongest edges of the network, where $w_l^{rank} \leq w_{l-1}^{rank}$ ($l = 1, \ldots, E_{>k}$) are the ordered weights on the edges.

McAuley *et al.* [74] examined the rich-club phenomenon across several layers of network connectivity including indirect edges of second and third neighbours of the original network. Higher layers of network connectivity (i.e. those of $2^{nd}$ and $3^{rd}$ neighbours) often revealed just the opposite rich-club behaviour as compared to the observation of rich-clubs of direct network edges. Examining the rich-club of interdependent networks a recent contribution [75] showed the existence of a "tricritical point" separating different behaviours as a response to node failures. Another recent finding showed that the observability of the network becomes maximal, when its hubs are dissociated from each other and do not form a rich-club [76].

*2.3 Network nestedness and its measures*

The nested property of a network (see Figure 2.) was first proposed, observed and defined in ecological systems [5,6,12], but recently it received much attention in the study of networks of the economy [22,77-80]. Ecological systems are usually described as incidence matrices (also called presence-absence matrices) defining bipartite networks [81]. Species-assemblages are nested if the species in species-poor sites are subsets of the assemblages of species in species-rich sites. In other words in a nested ecosystem the interactions of specialist species are usually a proper subset of the interactions of less specialist species [6].

Currently there are two widely used metrics for the characterization of nestedness: A.) the matrix temperature measure, *T*, defined by Atmar & Patterson [5] and B.) a nestedness measure based on overlap and decreasing fills, *NODF*, defined by Almelda-Neto *et al.* [82]. The matrix temperature, *T* quantifies whether the arrangement of 1's and 0's in the incidence matrix differ from the arrangement given by an isocline that describes a fully nested benchmark graph. The values of *T* are in the range from 0 to 100, and nestedness, *N*, is defined as

(11) $$N = 100 - T$$

where *N*=100 is the maximum nestedness level. Almeida-Neto *et al.* [83] pointed out some inconsistencies of the *T* matrix temperature measure, and in another paper [82] suggested a new metric, called *NODF*. The *NODF* metric is based on two properties: the decreasing fill and paired overlap. For a given m × n matrix let $N_i$ is the degree of node *i* (i.e. the sum of 1's of any row or column *i*). For any pair of rows *i* < *j* define $DF_{ij}$ = 100, if $N_i < N_j$ and $DF_{ij}$ = 0 otherwise. Let $DF_{kl}$ is similarly defined for any pair of columns *k* < *l*. Rows paired overlap is defined as $PO_{ij} = |N_i \cap N_j|/|N_i|$ and for columns *k* and *l*, $PO_{kl}$ be similarly defined. For any *i* < *j* row pair (and any *k* < l column pair), the degree of paired nestedness is defined as

(12) $$N_{ij} = \begin{cases} 0, & \text{if } DF_{ij} = 0 \\ PO_{ij}, & \text{if } DF_{ij} = 100 \end{cases}$$

The *NODF* measure of nestedness can be calculated by averaging all paired values of rows and columns:

(13) $$NODF = \frac{\sum_{i<j} N_{ij} + \sum_{k<\ell} N_{k\ell}}{\binom{n}{2} + \binom{m}{2}}.$$

One of the most important features of *NODF* is that it calculates the nestedness independently for rows and columns. Another important feature is the versatility of *NODF* enabling the evaluation of the nestedness of one or more columns (or rows) in relation to others [82].

Bastolla *et al.* [84] introduced an explicit definition of nestedness similar to that of *NODF*. However, it is very important to note that the measure of Bastolla *et al.* [79] is the only nestedness measure, which is directly linked to the dynamics (in particular: to the inter-species competition) of the complex system and thus to the development of the mutualistic plant/pollinator and plant/seed-dispersal networks.

Lee *et al.* [85], following [82], defined the nestedness of a unipartite network of n nodes with adjacency matrix $A = (a_{ij})$ as

$$S = \frac{1}{n(n-1)} \sum_{i=1}^{n} \sum_{j=1}^{n} \frac{\sum_{\ell=1}^{n} a_{i\ell} a_{j\ell}}{\min(N_i, N_j)}.$$

(14)

It is straightforward to extent *S* to bipartite networks. Equation (14) is also almost equivalent to equation (13) defining *NODF*, but it is easier to calculate it for a given matrix.

Recently, Podani & Schmera [86] proposed two other formulae for measuring nestedness: the "*percentage relativized nestedness*" (PRN), and the "*percentage relativized strict nestedness*" (PRSN). The formula of PRN satisfies the requirement that both similarity (overlap) and dissimilarity (the difference in the number of the two types of bipartite network nodes) should equally influence PRN. Let $a_{kl} = |N_k \cap N_\ell|$, the number of shared neighbours of nodes $k$ and $\ell$, $b_{k\ell} = |N_k \setminus N_\ell|$, the number of neighbours of node $k$ only, and similarly, $c_{k\ell} = |N_\ell \setminus N_k|$, the number of neighbours of node $\ell$ only. Then, let

$$\bar{N}_{rel} = \begin{cases} \frac{1}{\binom{n}{2}} \sum_{k<l} \frac{a_{kl} - |b_{kl} + c_{kl}|}{a_{kl} + b_{kl} + c_{kl}}, & \text{if } a_{k\ell} > 0 \\ 0, & \text{otherwise} \end{cases}$$

(15)

and let *PRN* be defined as $100 \bar{N}_{rel}$. *PRSN* is defined very similarly to *PRN*, but the condition $a_{k\ell} > 0$ is changed to $a_{k\ell} > 0$ and $b_{k\ell} \neq c_{k\ell}$. Using these notions above it can be obtained that

$$NODF_r = \begin{cases} \frac{100}{\binom{n}{2}} \sum_{k<l} \frac{a_{kl}}{a_{kl} + b_{kl}}, & \text{if } a_{k\ell} > 0 \\ 0, & \text{otherwise} \end{cases}$$

(16)

for rows (and similarly for columns). Podani & Schmera [87] underlined that *NODF* (in contrast to *PRN* or *PRSN*) depends on the ordering of columns in the data matrix. They advised the use of the expressions index, function or coefficient of nestedness instead of the metric [87]. In a recent publication, Ulrich & Almeida-Neto [88] warned that, despite the use of the concept of nestedness for more than five decades in ecology, there is still a large controversy regarding its precise meaning and applications. They noted that the *PRN* index includes tied ranks of nodes and counts them positively, while the *NODF* index penalizes tied ranks. Ulrich & Almeida-Neto [88] argued that extending nestedness to tied ranks would decrease the contribution of the key component of network asymmetry to nestedness.

Staniczenko *et al.* [89] recently proposed a new detection method that follows from the basic property of bipartite networks and shows how large dominant eigenvalues are associated with highly nested configurations.

Nestedness is in a complex relationship with other network measures. Nestedness usually increases with the number of interactions in the network [6]. Moreover, degree heterogeneity (the existence of hubs) has a very strong positive influence on nestedness [85,90]. If degree heterogeneity was discounted, nestedness was found to be correlated with degree disassortativity [90]. Nestedness in bipartite networks depends on the ratio of the number of nodes in the different classes of nodes of the bipartite network (species, colour classes, etc.), and nestedness becomes much larger in strongly heterogeneous scale-free networks [85]. At low connectivity, networks that are highly nested tend to be highly modular; the reverse is true at high connectivity [91]. Due to these effects, an extreme care must be exercised when comparing the nestedness of sparse and dense networks [89,92-95]. The use of various null-models [25,78,96] also has a paramount importance for the estimation of nestedness, since the distribution of values generated by null models also depend on the unique characteristics of each network. Hence, future work should be carried out in finding appropriate ways of comparing nestedness across networks.

*2.4 Bow-tie network structures*

The bow-tie structure refers to a core/periphery structure of directed networks (see Figure 1). Due to the directedness of the edges the bow-tie has a fan-in component of incoming edges (initiated at source-nodes) and a fan-out component of the outgoing edges (leading to sink-nodes). These two sides of the bow-tie surround the core, which is a highly intertwined giant component having nodes usually connected to both incoming and outgoing edges. The core of the bow-tie structure: A.) efficiently reduces the required number of nodes and edges to connect all source- and sink nodes; B.) decreases the effect of perturbations and noise; and C.) in case of biological networks, increases evolvability [7-9].

The BowTieBuilder algorithm of Supper *et al.* [97] gives a numerical score of "bow-tie-ness". BowTieBuilder searches the most probable pathway that connects the source- and sink-nodes of a potentially bow-tie structured network. The algorithm was originally defined and used for a signal transduction network, but it also can be applied to any directed, weighted network. In a general formalization $G = (V, E, w)$ is a directed graph, where each $e \in E$ link has the weight, $w_e \in [0, 1]$. The aim is to find a subgraph $G' \subseteq G$ that connects a set of source-nodes $S \subseteq V$ (in-nodes) to a set of sink-nodes $T \subseteq V$ (out-nodes or target nodes) ($T \cap S = 0$). The optimal solution of the problem is a subgraph $G'$ that has for every $s \in S$ and for every $t \in T$ at least one $(s, t)$-path, if such a path exists in $G$. The algorithm is a greedy approach to construct a pathway $P$ from a source-node to a sink-node, where the overall confidence of pathway $P$, $W_{prod}(P)$ given by

(17)
$$W_{prod}(P) = \prod_{e \in P \subseteq E} w_e$$

is maximal. BowTieBuilder favours pathways that are bow-tie structured. The bow-tie score of node $v$, $b(v)$ was defined to determine the core nodes of the network:

(18)
$$b(v) = \frac{|S_v||T_v|}{|S||T|}$$

where $|S_v|$ is the number of source nodes from which $v$ can be reached, $|T_v|$ is the number of target nodes that can be reached from $v$, $|S|$ and $|T|$ are the total number of source and target nodes, respectively. The bow-tie score is a centrality type measure [58].

Bow-ties lie in the middle of hierarchical and anti-hierarchical directed networks. "Hierarchical" and "anti-hierarchical" refer to tree-like, top-down and inverted tree-like, bottom-up network structures, respectively, as described by Corominas-Murtra *et al.* [98]. Bow-tie structures characterize the World Wide Web, the Internet, several manufacturing processes, the immune system, as well as metabolic, gene regulatory and signalling networks [7-9,97,99-102]. Similarly to the general core/periphery networks [31], rich-clubs [73], nested [91] and onion networks [33], bow-tie networks may also be modular [26,102].

*2.5 Onion networks*

A robust network should be resistant against both random failures (errors) and malicious attacks targeting its most important, topologically speaking, vital nodes. Scale-free networks are resistant against random failures, but are sensitive for targeted attacks [103]. Therefore the task may be formulated as an improvement of the remaining connectivity of scale-free networks after an attack with a minimal number of interventions concentrating to re-wiring instead of changing nodes. The seminal paper of Schneider *et al.* [10] used successive random edge-swaps, and found that the optimal network structure having the above robustness has an onion structure. (It is important to note that this complex type of connectivity-stability is not the same robustness as the network dynamics-related robustness defined by Kitano [8].) Schneider *et al.* [10] defined robustness as

(19)
$$R = \frac{1}{N} \sum_{Q=1}^{n} s(Q)$$

where $N$ is the number of nodes in the network and $s(Q)$ is the fraction of nodes in the largest connected component after removing $Q = qN$ nodes. The normalization factor $1/N$ ensures that the robustness of networks with different sizes can be compared. The range of possible $R$ values is between $1/N$ and $0.5$, where these limits correspond, respectively, to a star network and a fully connected graph. In the work of Schneider *et al.* [10] degrees were re-calculated after each attack to obtain networks that withstand an even more harmful attack strategy [10]. Onion networks are characterized by a core of highly connected nodes hierarchically surrounded by rings of nodes with decreasing degree (see Figure 2). The onion structure implies that almost every node remains connected after removing the most important nodes of the network, the hubs in the core [10,15,104].

Wu & Holme [15] provided a generative algorithm to obtain onion networks, which had linear computational complexity instead of the cubic complexity of the algorithm of Schneider *et al.* [10]. First, a set of $N$ random numbers $\{k_i\}$ was generated drawn from a distribution $P(k) \sim k^{-\gamma}$. These numbers represented the degrees of the $N$ nodes in the network. Each node $i$ was then assigned a layer index $s_i$ according to its $k_i$ value. Nodes were ranked by increasing degree. The layer index for nodes with lowest degree was set as 0, while the layer index for the node-sets with increasingly larger degrees was increased to 1 and higher numbers until all nodes have been assigned a layer index, $s_i$. Then half-edges were connected by selecting a pair of nodes at random, and joining these with a probability dependent on the layer difference of the two nodes according to the formula

$$P_{ij} = \frac{1}{1+\alpha|\Delta_{ij}|}$$

(20)

where $\Delta_{ij} = s_i - s_j$ is the difference in layer index between nodes $i$ and $j$, and $\alpha$ is a control parameter being optimal (from the point of onion network structure formation) at intermediate values between zero and infinite. There is a fraction of half-edges (usually in the range of 1 to 2 %), which can not be paired, and require an additional reshuffle procedure. Importantly, the onion networks created by the procedure of Wu & Holme [15] were very close to the optimal networks of Schneider *et al.* [10].

Onion networks lie in the boundary of assortative and disassortative networks, as well as fully connected graphs and star networks providing stability of network connectivity against targeted and random attacks, respectively [10,15,16,105,106].

The recent work of Louzada *et al.* [33] offered a faster rewiring process creating alternative connections between parts of the network that would otherwise be split upon the failure of a hub without degree re-calculation. Their method preserved the modularity of the original network and resulted in modular onion structures. The method combined the concepts of the above, connectivity-related robustness and the Harary index [107] or network efficiency [108], $E$, which was defined as

(21)
$$E = \sum_{\substack{i,j=1 \\ i \neq j}}^{n} \frac{1}{l_{ij}}$$

where $l_{ij}$ denotes the shortest path length between nodes $i$ and $j$ [108]. Using this formula, Louzada et al. [33] defined the "Integral efficiency of a network", IE, as

(22)
$$IE = \frac{1}{N} \sum_{Q=1}^{n} E(Q)$$

where $E(Q)$ is the efficiency of the network after removing $Q$ nodes. They presented another optimization method to increasing $R$ and $IE$ by using an exponential function for the acceptance probability of edges swapping. The recent results of Tanizawa et al. [109] based on analytical considerations, also confirmed that an onion-like network structure is a nearly optimal candidate against removal of a random or a high degree node.

So far no real world examples have been found of onion networks. The discovery of these naturally developed robust networks may be hindered by conceptual elements, since onion networks are considered as rewired real world networks after human intervention to enable them to withstand both random failures (errors) and targeted attacks. What if this strategy has already been applied in some types of networks? The wheel networks of criminal organizations (such as Colombian drug trafficking networks) described by Kenney [110] have a dense core and a ring of nodes connected to the core, and can certainly be considered as single-ring onion networks. In the network of the 19 hijackers and 18 covert conspirators of the September 2001 attacks a ring-like network segment is constructed by covert conspirators improving communication and preserving hijackers' small visibility and exposure [111,112]. It is a question of future studies, whether other criminal networks developed an onion structure. If many criminal networks have an onion structure the additional questions arise, if these networks usually have a single core with a single ring, have multiple rings, or display multiple cores. Existing data and assumptions support the multiple core structure, called cluster-and-bridge organization [113,114]. It is also an interesting question, whether the connectivity-related network robustness of onion networks increases further, if the core is not in their centre but on their side. Network connectivity-related robustness tests of these networks may include a concept of "indirect attack", where the attack is channelled by a neighbour of the targeted node (who was arrested by the authorities and reveals the identity of his/her neighbours in the network).

*2.6 Similarities and differences of core/periphery network structures*

Four of the core/periphery type networks, rich-clubs, nested networks, bow-ties and "traditional" core/periphery networks are rather similar to each other in the sense, that all of them have a highly connected core (often containing hubs) and peripheral nodes, which are preferentially connected to the core, but usually are not connected to each other. As already Borgatti & Everett [1] noted, "all actors in a core are necessarily highly central (…). However, the converse is not true, as not every set of central actors forms a core." While core/periphery coefficients, rich-club coefficients and nestedness indices in principle can be extended to weighted and directed networks, bow-tie structures are more restricted, since they characterize only directed networks. (Importantly, core-periphery indices of weighted and directed networks will show a rather different picture than those of unweighted and undirected networks.) Nestedness indices mostly characterize bipartite networks. The onion network is different from the other four in the sense, that in onion networks peripheral nodes are also connected to each other – albeit preserving their connections to the network core (see Figure 2). A general comparison of various core indices is clearly missing, and will be a crucially important task of future studies.

Core/periphery structures and network modules are two representations of the development of dense network structures. Both core/periphery structures and network modules are meso-scale network components, and display a high level of complexity. Core/periphery structures lie in the middle of several extreme properties, such as random/condensed structures, clique/star configurations, network symmetry/asymmetry, network assortativity/disassortativity, as well as network hierarchy/anti-hierarchy. These properties of high complexity greatly contribute to their high robustness in the network dynamics sense [8,10,15,16,98,105,106,109].

Cores are different from module centres, since A.) they are surrounded by peripheral nodes, which are not connected to each other; and B.) usual core/periphery structures contain only one core, while a network usually consists of multiple modules. However, we have to note that there is no clear discrimination between network cores and network modules, since in onion networks [10,15,16,33] peripheral nodes do connect each other, and multiple network cores were also described [31,33].

## 3. Dynamics of core/periphery networks

In the previous section we have focused on the structural properties of core/periphery networks. In this section, we will discuss the dynamics of different core/periphery networks including their development upon changes of environmental conditions, such as available resources to build and maintain network connections, as well as constraints, restricting or destroying network edges or nodes.

Core/periphery structures may develop, if system resources become low or environmental stress increases. Both conditions may lead to the development of more condensed network structures, such as the segregation of a network core, as well as the formation of well-separated network modules [17-22,108,115-117]. Importantly, network analytical considerations allow the development of network cores much better, if the "length" of edges is also considered as part of their costs [19]. The increased contribution of the costs of edge length may be a reason why flow-type networks (such as metabolic networks, signalling networks, the Autonomous Systems of the Internet, etc.) often develop a more characteristic core/periphery structure than association-type networks (such as protein-protein interaction networks, social networks, etc.) [2]. Importantly, this network classification is similar to that given by Guimera *et al.* [118] assessing the topology of hubs and non-hub nodes in modular networks.

The development of cooperation may also lead to the segregation of a cooperating core of social networks, pushing out defectors to the network periphery [119]. In agreement with this observation in social networks, nested plant/pollinator and plant/seed disperser ecological networks were shown to reduce effective inter-specific competition [84]. Yook & Meyer-Ortmanns [120] showed that synchronization of Rössler-type oscillators can be achieved only, if you make shortcuts of a Cayley-tree between the outer nodes and the central nodes and not between the outer nodes to each other. Additionally, a rich-club of neuronal networks was shown to induce the synchronization of neuronal oscillation patterns, which enhances further the generality of cooperation-supporting, competition-minimizing role of core/periphery structures [121,122].

Currently we do not have a clear understanding of the environmental and network structure conditions regulating the number of developing network cores (i.e. the development of a single "traditional" network core, or a multi-modular structure). Importantly, in the long-term, core-periphery structures are derived from evolutionary changes. However, core-periphery structures may also abruptly develop or transform. The exact conditions governing these short-term changes are rather unclear. Conditions regulating the size of the developing core (i.e. the fuzziness of the core structure) are also not entirely clear. Smaller and tighter core may develop, when resources become poorer. In extreme pauperization of resources and/or during extremely large stress the core may condense to single hub developing star-network [19], which is a well-known network structure at small resources and/or large stress [17-19]. Alternatively, the core/periphery structure may be transformed to a chain structure, which happened with the onion-type Colombian drug trafficking networks under severe law enforcement attacks between 1989 and 1996 [110].

A smaller core may enable a tighter controllability of the network (in the sense of maximally achievable control), but may also lead to a smaller flexibility and adaptability [9,26,27,122-124]. It seems that extreme conditions may induce more specialized system behaviour with a smaller core and tighter control. In other words: large core size may increase the plasticity of system behaviour shifting it closer to the state "at the edge of chaos". Smaller cores, which may characterize stressful conditions, may ensure more focused, more efficient, more rigid system behaviour. Conditions leading to the enlargement of the core and even to its "dissolution" may include higher resources and less environmental stimuli. However, currently we do not exactly know, in which cases core-segregation becomes reversible.

Decreasing core-size leads to the emergence of signalling bottlenecks in signalling networks, which may critically reduce signalling capacity especially in the presence of noise [125]. Information handling capacity is presumed to be minimally sufficient [126], thus core-reduction may lead to a reduction of system-responsiveness to external stimuli. In agreement with this assumption, stress leads to a general reduction of system responses [18]. Network cores may participate in the formation of switch-type responses after surpassing a sensitivity threshold. Currently we do not know how core/periphery structures may regulate amplification, filtering, threshold development and sensitivity of signalling networks.

The development of network core increases network robustness and stability [8] in a large variety of real world networks. This is mainly due to the rich connection structure of the core allowing a high number of degenerate processes, ensuring cooperation and providing multiple options of network flow re-channelling, when it is needed. Importantly, core processes enable a coordinated response of various stimuli. The core usually has much less fluctuations than the periphery, and has much more constraints, therefore changes (evolves) slowly. The integrative function of network cores is an important step in the development of a large variety of complex systems [8,9,12,14,21,78,79,84,123,124,127-130].

## 4. Function of core/periphery network structures in different types of real world networks

In this section, we will discuss the functions of core/periphery network structures in real world molecular networks (including protein structure networks, protein-protein interaction networks – interactomes –, metabolic networks, signalling networks, gene regulatory networks and chromatin networks), cellular networks (including the immune and the nervous systems), ecological networks, social networks and networks of the economy. The functional consequences will summarize both the structure and dynamics of all types of core/periphery networks we described in the previous chapters (i.e. traditional core/periphery structures, rich-clubs, nested networks and bow-tie networks). Onion networks will not be included, since currently no real world networks are known which fulfils their definition criteria.

*4.1 Protein structure networks*

In protein structure networks nodes usually represent the amino acid side chains (in some works networks of individual atoms are also used). Edges represent the noncovalent interaction between amino acids. Edges of unweighted protein structure networks connect amino acids having a distance below a cut-off distance, which is usually between 0.4 to 0.85 nm. In weighted protein structure networks, the edge weight is usually inversely proportional to the distance between the two amino acid side-chains [131]. Structures of globular proteins are naturally organized to core/periphery networks: hydrophobic amino acids form the core, while hydrophilic and charged amino acids are on the periphery enabling a contact structure with the surrounding water. Core residues evolve slowly, while peripheral residues show a much faster evolution. Importantly, if peripheral residues are involved in protein-protein contacts, their evolutionary rate is slower, which "freezes" the evolution of the core. This makes the interactome hubs especially conserved [128,129]. Though physical constraints do not allow the development of mega-hubs in proteins, protein structure networks possess a rich-club structure with the exception of membrane proteins, where hubs form disconnected, multiple clusters [132]. The rich-club structure is especially true for the hydrophobic core of proteins [133]. However, rigorous studies of core/periphery structures of protein structure networks and their possible reorganization during allosteric signalling are currently missing.

*4.2 Molecular networks: interactomes, metabolic and signalling networks*

Nodes of protein-protein interaction networks (interactomes) represent proteins and the edges represent physical interactions between them. Interactomes are probability-type networks; that is, their edge weights reflect the probability of the actual interaction [131]. Interactomes are, therefore, usually not directed. However, recently directions of protein-protein interactions could be defined by the extension of the directions observed in signalling networks using learning algorithms [134]. Interactomes have a clear core/periphery structure. Core proteins are rather conserved, many times essential, and mainly perform general functions, which are independent of their tissue or organ distribution. Peripheral proteins tend to be localized towards the physical periphery of the cell, and mainly perform organ-specific functions [130,135,136]. A recent study showed that – in contrast to other age-related gene classes, including longevity- and disease-associated genes, as well as genes undergoing age-associated changes in gene expression – age-associated changes in DNA methylation patterns occurs preferentially in genes that occupy peripheral network positions of exceptionally low connectivity [137]. Hubs of protein-protein interaction networks mostly reside in different communities and do not form a direct rich-club with each other [4,71]. The lack of rich-club structure prevents inactivation cascades caused by free-radical damage, and may increase the flexibility of responses. However, a second-order rich-club of the interactome has been observed, when not the direct neighbours but the second neighbours were analyzed. This second order rich-club reinforces the organization of the interactome core, and may ensure that key proteins act in an integrated manner [74].

In metabolic networks, major metabolites (nodes) are connected by enzyme reactions (edges related to the corresponding enzyme/s/), which transform them to each other [131]. Bow-tie network structures were first detected in metabolic networks [7]. Later studies showed that the relative core-size of cellular metabolism varies from organism to organism and may be modular. Cores extend to a larger segment of the small, specialized metabolism of organisms living in constant environmental conditions (such as symbiont bacteria), while a smaller ratio of core reactions represents organisms experiencing a large variability of environmental changes requiring a large variability of "fan-in" reactions (such as free living bacteria). Enzymes catalyzing the core reactions had extended mRNA half-lives, and had a considerably higher evolutionary conservation – in agreement with similar observations in interactomes [130,138-140]. Bow-tie structures reveal vulnerable connections, especially in their areas connecting the core and the fan-in/fan-out components, which may be used for drug targeting [130,131,139].

The other intensively studied directed cellular networks, signalling networks, also show a bow-tie structure. Signalling networks represent the segment of the interactome, genes and micro-RNA-s involved in cellular signalling [131]. Plasma membrane receptors and other proteins at the beginning of signalling cascades usually represent the fan-in component of the bow-tie. Transcription factors are often in the bow-tie core, while the induced genes and their regulatory microRNA-s are in the fan-out side. Similarly to metabolic networks, the core of signalling bow-ties may also be modular and even pathway-specific [97,141-144]. In agreement with the general assumption on the integration provided by core/periphery structures, bow-tie structured signalling was shown to produce much more integrative responses than segments of signalling network lacking this organization [145].

Gene regulatory networks are the downstream parts of signalling networks, which contain transcription factors, their DNA-binding sites at the genes they regulate, the genes themselves and the transcribed messenger RNAs, as well as microRNA-s regulating gene expression by binding to complementary sequences on target messenger RNAs [131]. Gene regulatory networks themselves have a core/periphery structure. Core master transcription factors are vital for survival and are almost continuously active. The core/periphery structure of gene regulatory networks reduces transcriptional noise, and provides a key mechanism of signal integration [21]. Here again (as generally with the extended signalling networks), a bow-tie structure may be observed, which may contain a single or a modular core. The core becomes condensed in free-living bacteria as compared to e.g. symbionts – in agreement with the similar observations studying metabolic networks (cf. references [32], [102] and [130]). Chromatin-interaction networks (i.e. distant segments of DNA, often in different chromosomes providing nodes – connected by the large protein complex of transcriptionally active RNA polymerase II providing edges) revealed a rich-club structure, which was enriched in essential processes, and may provide a structural and functional robustness, integration and cooperation of transcriptional processes [127]. The core/periphery organization of gene-regulatory and chromatin-interaction networks may be related to the recently discovered super-enhancers determining cell differentiation and disease conditions, such as cancer [146,147].

*4.3 Cellular networks: neurons, immune system*

Individual cells of the immune system or neuronal cells have a high specialization. Moreover, they are mobile (immune cells), or may develop highly mobile, extremely long extensions (neuronal cells). Therefore, they are able to form much more complex networks than other cells, where functions of high complexity, such as the discrimination from self to non-self, or human consciousness, may emerge [18]. The immune system was shown to have a bow-tie core/periphery structure centralized to $CD^+$ naïve T-cells [101].

Networks of neurons of cat cerebral cortex and human brain showed a rich-club organization, providing A.) a backbone of controlled synchronization of neuronal oscillations necessary to higher cognitive tasks; B.) its easy regulation by changing the oscillatory pattern of a single hub, as well as C.) a central, high-cost (40% of total), high-capacity backbone for global brain communication. The rich-club of human brain resembles to a bow-tie structure in the sense, that many dynamic pathways are first fed into, then traversed by and finally exited from the rich-club structure. Task-related activation patterns often include an activity-shift from peripheral neurons to their connected rich-club neurons [121,122,148-150]. Interestingly, a transient increase of rich-club-like properties was observed in near-death brains of rats during cardiac arrest demonstrating neural correlates of paradoxically heightened conscious processing in near-death brains [151]. On the contrary, in deep-sleep a breakdown of long-range temporal dependence of default mode and attention networks was observed [152], which may indicate a transient disintegration (and consequent re-organization) of the rich-club structure.

In agreement with the previous observations the recent paper of Bassett *et al.* [153] demonstrated that the learning process of human brain can be described by the presence of a relatively stiff core of primary sensorimotor and visual regions, whose connectivity changes little in time, and by a flexible periphery of multimodal association regions, whose connectivity changes frequently. The separation between core and periphery is changing with the duration of task practice and, importantly, is a good predictor of individual differences in learning success. Moreover, the geometric core of strongly connected regions tends to coincide with the stiff temporal core. Thus, the core/periphery organization of the human brain (both in its structure and dynamics) plays a major role in our complex, goal oriented behaviour [153].

*4.4 Ecological networks*

Ecological networks often display a nested structure, where specialists interact with generalists, while generalists interact with each other and with specialists [5,6]. The ecological concept of nestedness was published first in [154], and it was formally defined by Atmar & Patterson [5]. The stability of an ecological network strongly depends on the pattern of the shared interactions, and not only on pairwise interactions between species. Bascompte *et al.* [6] studied 52 mutualistic ecological networks, and showed that they are highly nested relative to the null-model.

Burgos *et al.* [155] investigated the relationship between the nestedness and robustness of mutualistic systems and established that the nested pattern is best possible one for enhancing robustness, provided that the least linked species have the larger probability to extinction. Suweis *et al.* [156] showed that nested interaction networks could emerge as a consequence of an optimization principle aimed at maximizing the species abundance in mutualistic communities. Particularly, an increase in abundance of a given species results in a corresponding increase in the total number of individuals in the community, and also an increase in the nestedness of their interaction matrix. On the contrary, the abundance of the rarest species is linked directly to the resilience of the community. Bastolla *et al.* [84] showed that the competition between different species is decreasing, and the number of coexisting species is increasing with increasing nestedness.

Saavedra *et al.* [78] investigated several highly nested flowering plant/insect pollinator networks, and showed that nodes contribute very heterogeneously to the nested architecture of the network (the individual contribution to nestedness for each node was calculated by the node-level metric, defined as *z*-score of the network nestedness by using appropriately randomized null-models). From simulations, Saavedra *et al.* [78] observed that the removal of a strong contributor to the overall nestedness decreases the overall persistence of the network more than the removal of a weak contributor, where persistence of the network referred to the proportion of survivors (which was calculated using the dynamical equations introduced by Bastolla *et al.* [84]). Strong contributors to the nestedness of the network (and its persistence) were shown to be the most vulnerable one to extinction [78]. As their recent work showed [157], species tolerance is extremely sensitive to the direction of change in the strength of mutualistic interaction, as well as to the observed mutualistic trade-offs between the number of partners and the strength of the interactions. Generalists and strong contributors to nestedness are the most tolerant to an increase in the strength of mutualistic interaction.

*4.5 Social networks*

Social interactions can be represented as networks, where nodes represent persons and edges stand for their social relationships. One of the first core/periphery network examples of Borgatti & Everett [1] was the social network of 20 monkeys, where the 5 males dominated the core, and the 15 females formed a periphery. Most of social communities form an elite group, i.e. a small core of well-connected network members, who are also connected to each other. Members of the "elite" are often those, who established a new group. Newcomers at the network periphery are often connected to other members of the periphery, while establishing connections with the "elite" later. Connections to 2 "elite"-members, who know each-other (where these triangles are called as simmelian ties) were shown to provide the best help to become a member of the "elite" later [1,158,159]. In general agreement with these data Banos *et al.* [160] showed that as a Twitter social network got „older" (i.e. more organized) its communication became more centralized to its core. In agreement with other social network studies the core is formed by the elite of users that grew more prominent during the exchange of Twitter information. On a similar topic of the emergence of social networks, Guimera *et al.* [161] proposed a model for team assembly based on three parameters: team size, the fraction of newcomers in new productions, and the tendency of incumbents to repeat previous collaborations. The model indicates that the emergence of a large connected community of practitioners can be described as a phase transition. Further, depending on the fraction of incumbents and newcomers and the propensity for repeated relationships, they found that assembly mechanisms determine both the structure of the collaboration network and team performance for both artistic and scientific fields.

The rich-club organization of social hubs was first observed studying intensively collaborating scientists [4]. Second and third neighbour rich-clubs are not preserved in scientific collaboration networks, reflecting the fact that scientists are usually collaborating within their scientific areas [74,161,162]. In experimental social dilemma games involving human subjects Ahn *et al.* [119] showed that cooperators form a network core of interacting hubs, and expel defectors to the network periphery. Rich-club and nested network structures also showed "super-cooperation" [163]. Importantly, besides the general interconnectedness of hubs, mega-hubs ("mega-stars") of social networks often tend to avoid collaboration, and form subgroups of potential rivalry in the network core [71-73,164-167]. On the contrary, Twitter shows an especially high concentration of inter-connected mega-hubs, where the top 1,300 PageRank users of Twitter may impose their opinion to the whole network of more than 41 million users [168,169]. Reasons for the development of mega-hub avoidance may lie in the cost/risk/benefit ratios, but are not clear [170,171], and require further investigations.

*4.6 Networks of the economy*

The core/periphery structure is especially characteristic of the market-related interactions between persons or social groups, which can be perceived as networks of the economy. Networks of the industry, as well as bipartite networks of industrial firms and their locations show a core/periphery structure and a high nestedness [77-79,172-186]. Moreover, nestedness of the network of industrial compounds and their locations was nearly constant over many years. This "conservation law of industrial nestedness" can be used to predict the evolution of industrial systems [79]. Removal of a strong contributor to the nestedness of the New York garment industry network decreased the overall persistence of the network more than the removal of a weak contributor, where persistence of the network referred to the proportion of survivors [78,177]. This finding reinforced the view that nested structure is not only evolving and maintained but also important for the stability of networks in the economy.

Studying the evolution of Research and Development (R&D) industrial networks between 1986 and 2009 Tomasello *et al.* [178] found that R&D networks show an increased core/periphery coefficient [2] and nestedness [96] at the "golden age" of R&D industry between 1990 and 1997. Nestedness values were high, and increased further, when the partnership was more costly. The only notable exception was the pharmaceutical R&D industry, where the most pronounced core/periphery structure took place between 2002 and 2005 [178]. This finding is in agreement with the fact that the pharmaceutical industry started a massive outsourcing of R&D capacities around 2003 to 2005 [131]. Uzzi & Spiro [166,167] examined the evolution of the network of creative artists who made Broadway musicals from the inception of the industry in the late 1900s to the present day, and found that nestedness increased between shocks – upheaval events such as world wars, depressions, and infectious diseases – that restructured the network.

A core/periphery organization with a bow-tie structure was observed in the control network of trans-national corporations. Network control was found to be even more unequally distributed than wealth. A large portion of control flowed to a small tightly-knit core of financial institutions. Top ranked actors held a control ten times bigger, than what could be expected from their wealth. About ¾ of ownership of firms in the core remained in the hands of firms of the core [123]. This structure was shown to arise by a "rich get richer" mechanism without the need of an explicit underlying strategy of the companies [124]. However, such a level of centralization may raise questions on market competition and stability.

The world trade can be described by a bipartite graph, where nodes represent either countries, or products, and weighted edges between a country and a product represent their ratio related to the total amount of the product imported or exported. This World Trade Web is highly nested [80]. Such nestedness remained constant between 1985 and 2009 [79], and most probably contributes a lot to the stability of the world trade. (The stability of the international trade network – as opposed to e.g. the Dutch interbank network – was recently confirmed by a different method [179].) Importantly, several key players of the World Trade Web, such as China, France, Germany, Japan, the UK and the USA, shared on average less weight among them than anticipated from the random situation [75]. This "aversion" of key players is similar to the competitive dissociation of key members of the social elite [71-75,164,165] mentioned in the preceding section. Interestingly, distances of countries did not play a crucial role shaping the World Trade Web [180].

*4.7 Engineered networks: the Internet, the World Wide Web, transportation networks and power grids*

After the "products" of social interactions related to the market, last, but not least, we summarize the core/periphery structures of man-made objects. The Internet was one of the first engineered systems, where a core/periphery structure was observed. Importantly, the bow-tie type core structure of the Internet is relevant to the "richest nodes" of the Autonomous Systems (AS) level, while the high degree hubs, which provide Internet connectivity to local regions, are not directly inter-connected, thus do not form a *bona fide* rich-club [3,4,9,71,100,101,181]. Importantly, the Autonomous Systems core of the Internet remained stable despite the rapid development of the system [51], which is a similar phenomenon to the "conservation of nestedness" of the networks related to the economy [79] and extends the examples showing the core-dependent stabilization of complex systems.

The World Wide Web is not really an engineered, but rather a social network-derived system. Not surprisingly, its core/periphery character is similar to that of social networks in the sense that high in-degree web-pages purposefully avoid sharing links to other high in-degree web-pages (possibly decreasing the success of their competitors) [165].

Air transportation networks show a rich-club ordering, i.e. traffic is heavier among large airports than expected by chance [72]. Still, a competition between the largest US airports can be observed from their weighted network structure [73].

Power-grids show an unchanging rich-club phenomenon all in their first, second and third neighbour connection structures, which may indicate the network connectivity-related robustness of this network against blackouts, since several neighbouring hubs would be able to aid a faulty hub in case of an emergency [74].

**5. Conclusions and perspectives**

We conclude our review by summarizing a few general remarks and open questions for further research in the field of core/periphery networks.

Currently there is no clear discrimination between network cores and network modules. A core may be a "*global core*", if it is the only core in the network. Cores of modular structures can be classified as "*local cores*". A local core may also be a global core, if it is the only local core, which occupies a central position in the network, and all other local cores are peripheral. Such a situation occurs if the network has a hierarchical modular structure with a single module being at the top of the hierarchy. It is important to note that in real world networks a *bona fide* "global" core does not exist, because all real world networks can be extended beyond their "ends" – since they are connected to other real world networks – therefore the global nature of their core remains always relative.

Both the discrimination between network cores and network modules, as well as the discrimination between global and local cores require an appropriate mathematical formalism, and further work. Spectral scaling [23], which was used to discriminate between four topological network classes including modular and core/periphery networks [24] may give further insights to these topological differences. The novel concept of Bassett et al. [153] on discriminating core and periphery based on node/edge dynamics, where the core is rigid, and the periphery is flexible [182], also opens other new areas of discussion and discovery.

Core/periphery structures lie in the middle of several extreme properties, such as random/condensed structures, clique/star configurations, network symmetry/asymmetry, network assortativity/disassortativity, as well as network hierarchy/anti-hierarchy. These properties of high complexity greatly contribute to the high network dynamics-related robustness of core/periphery systems [8,10,15,16,98,105,106,109]. The robustness and stability of core structures is mainly due to their rich connection structure allowing a high number of degenerate processes, ensuring cooperation and providing multiple options of network flow re-channelling, when it is needed. Importantly, core processes enable a coordinated response of various stimuli. The core usually has much less fluctuations than the periphery, and has much more constraints, therefore changes (evolves) slowly. The integrative function of network cores is an important step in the development of a large variety of complex systems [8,9,12,14,21,78,79,84,123,124,127-131].

Cooperation is a key element of the efficiency of core/periphery structures ensuring integration, stabilization and evolvability at the same time at the systems level. Real world social dilemma games result in the evolution of a cooperating core [119,177]. Nested plant/pollinator and plant/seed disperser ecological networks reduce effective interspecific competition [84]. The rich-club of neuronal networks induces the synchronization of neuronal oscillation patterns [121,122]. Core/periphery networks are conserved in industrial networks [79] and in the Internet [51] and coincide with the "golden age" of R&D industry [178]. All these examples show the importance of core/periphery structures in the stabilization of fast-growing, fast-changing complex systems.

From the concluding remarks above it is quite clear that there is a lot of remaining work to clarify the structure, dynamics and function of core/periphery networks. Here we list a few of the open questions.

- What is a general definition of core/periphery networks and how can they be discriminated from modular networks? What other type of core/periphery definitions exist, which are based on node and edge dynamics defining the core as rigid, unchanged, while the periphery as flexible regions, respectively [153,182]?
- How can we develop a more detailed comparison and discrimination of edge-cores and node-cores – especially in weighted and/or directed networks?
- Since the distribution of values generated by null models also depend on the unique characteristics of each network, how can future work find appropriate ways of comparing core-periphery structures across networks using appropriately selected null-models?
- What tests are needed to develop a general comparison of various core indices (including rich-club coefficients, bow-tie scores, nestedness measures and onionness indices besides "traditional" core scores)? This will be a crucially important task of future studies including weighted and/or directed networks. Such a study must involve the comparison of null-models [1,2,4-6,25] and suggestions for their appropriate selection.
- What are the discriminatory features of the environmental and network structure conditions leading to the short-term condensation/dissolution of a global core? Where "short-term" discriminates these changes from the long-term, evolutionary time-scale.
- How and when multiple local cores are formed, transformed to a single global core or become dissolved in a short-term process?
- How do the absolute and relative sizes of the core(s) change, and what are the functional consequences of short-term core-expansion/repulsion?
- What is the similarity of the various core-measures, if compared in a variety of different real world networks in a rigorous and systematic manner? Which core-measures are

- more useful in different types of real world networks?
- What is the real-world equivalent of onion-networks, which are the only core-periphery networks that were discovered by an optimization procedure and not by the analysis of real world networks?
- What are the properties of core/periphery structures of protein structure networks and what are the dynamic changes of protein structure cores and peripheries during allosteric signalling?
- How can core/periphery structures regulate signal amplification, filtering, threshold development, signalling capacity and sensitivity?
- To what extent the network core and periphery corresponds to the physical location of network nodes in real world networks embedded in real space?
- Which are the specific conditions (thresholds) when, why and how hub-hub association becomes hub-hub avoidance (e.g. in social networks, and their "products", such as the World Wide Web, World Trade Web, etc.)?

As the above open questions clearly demonstrate, despite of several decades of research interest, studies on core/periphery networks still hold a lot of promises and future discoveries. The authors of this review would like to encourage their colleagues to enrich further this exciting field.

**Funding**

Work in the authors' groups was supported by the Hungarian National Science Foundation [OTKA-K83314 to P.C.], by the National Natural Science Foundation of China [11131009 and 91330114 to L.W.], by the Northwestern University Institute on Complex Systems (NICO) and by the Army Research Laboratory [Cooperative Agreement Number W911NF-09-2-0053 and DARPA BAA-11-64 Social Media in Strategic Communication to B.U.]. The views and conclusions contained in this document are those of the authors and should not be interpreted as representing the official policies, either expressed or implied, of the Army Research Laboratory or the U.S. Government. P.C. is thankful for the Chinese Academy of Sciences Visiting Professorship for Senior International Scientists allowing the completion of this manuscript.

**Acknowledgement**

Authors are thankful to Ernesto Estrada for encouraging the writing of this review. P.C. is grateful to Serguei Saavedra (Integrative Ecology Group, Estación Biológica de Doñana, CSIC, Sevilla, Spain) and István A. Kovács (Institute for Solid State Physics and Optics, Jenő Wigner Research Centre, Hungarian Academy of Sciences & University of Szeged, Hungary) for critical reading of the manuscript.

**Table 1. Definition of (locally) dense network structures**

| Name of dense network structure | Definition | References |
|---|---|---|
| **Clique** | a complete subgraph of size $k$, where complete means that any two of the $k$ elements are connected with each other | [36,37] |
| ***k*-clan** | a maximal connected subgraph having a subgraph-diameter $\leq k$, where the subgraph-diameter is the maximal number of links amongst the shortest paths *inside* the subgraph connecting any two elements of the subgraph | [37-39] |
| ***k*-club** | a connected subgraph, where the distance between elements of the subgraph $\leq k$, and where no further elements can be added that have a distance $\leq k$ from all the existing elements of the subgraph | [37-39] |
| ***k*-clique** | a maximal connected subgraph having a diameter $\leq k$, where the diameter is the maximal number of links amongst the shortest paths (including those *outside* the subgraph), which connect any two elements of the subgraph | [37-40] |
| ***k*-clique community** | a union of all cliques with $k$ elements that can be reached from each other through a series of adjacent cliques with $k$ elements, where two adjacent cliques with $k$ elements share $k$-1 elements *(please note that in this definition the term k-clique is also often used, which means a clique with k elements, and not the k-clique as defined in this set of definitions; the definition may be extended to include variable overlap between cliques)* | [41,42] |
| ***k*-component** | a maximal connected subgraph, where all possible partitions of the subgraph must cut at least $k$ edges | [43] |
| ***k*-plex** | a maximal connected subgraph, where each of the $n$ elements of the subgraph is linked to at least $n$-$k$ other elements in the same subgraph | [37,44] |
| **strong LS-set** | a maximal connected subgraph, where each subset of elements of the subgraph (including the individual elements themselves) have more connections with other elements of the subgraph than with elements outside the subgraph | [37,45] |
| **LS-set** | a maximal connected subgraph, where each element of the subgraph has more connections with other elements of the subgraph than with elements outside of the subgraph | [37,45,46] |
| **lambda-set** | a maximal connected subgraph, where each element of the subgraph has a larger element-connectivity with other elements of the subgraph than with elements outside of the subgraph (where element- | [37,47] |

| | connectivity means the minimum number of elements that must be removed from the network in order to leave no path between the two elements) | |
|---|---|---|
| **weak (modified) LS-set** | a maximal connected subgraph, where the sum of the inter-modular links of the subgraph is smaller than the sum of the intra-modular edges | [37,45] |
| ***k*-truss or *k*-dense subgraph** | the largest subgraph, where every edge is contained in at least (*k*–2) triangles within the subgraph | [48-51] |
| ***k*-core** | a maximal connected subgraph, where the elements of the subgraph are connected to at least *k* other elements of the same subgraph; alternatively: the union of all *k*-shells with indices greater or equal *k*, where the *k*-shell is defined as the set of consecutively removed nodes and belonging links having a degree ≤ *k* | [37,45,52] |

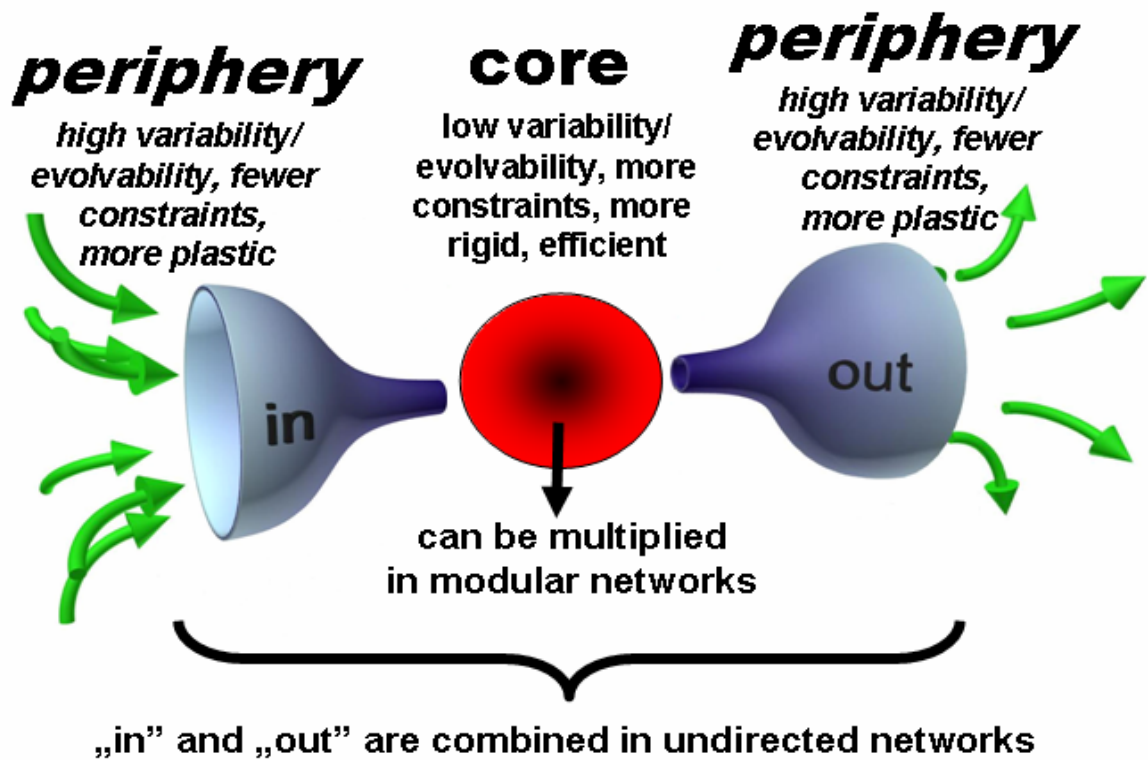

FIG. 1. General features of core/periphery network structures shown by the example of the bow-tie architecture of directed networks. The "in" and "out" components of network periphery refer to the fan-in and fan-out segments of bow-tie networks containing source and sink nodes, respectively. These segments of network periphery are combined in undirected networks. The network periphery has higher variability, dynamics and evolvability, has fewer constraints, and is more plastic than the core. Network cores facilitate system robustness helping the adaptation to large fluctuations of the environment, as well as to noise of intrinsic processes. The network core can be regarded as a highly degenerate segment of the complex system, where the densely intertwined pathways can substitute and/or support each other. The network core has lower variability, dynamics and evolvability, and is more rigid and more efficient than the periphery. Core structures may be multiplied in modular networks. Adapted from Tieri *et al.* [13].

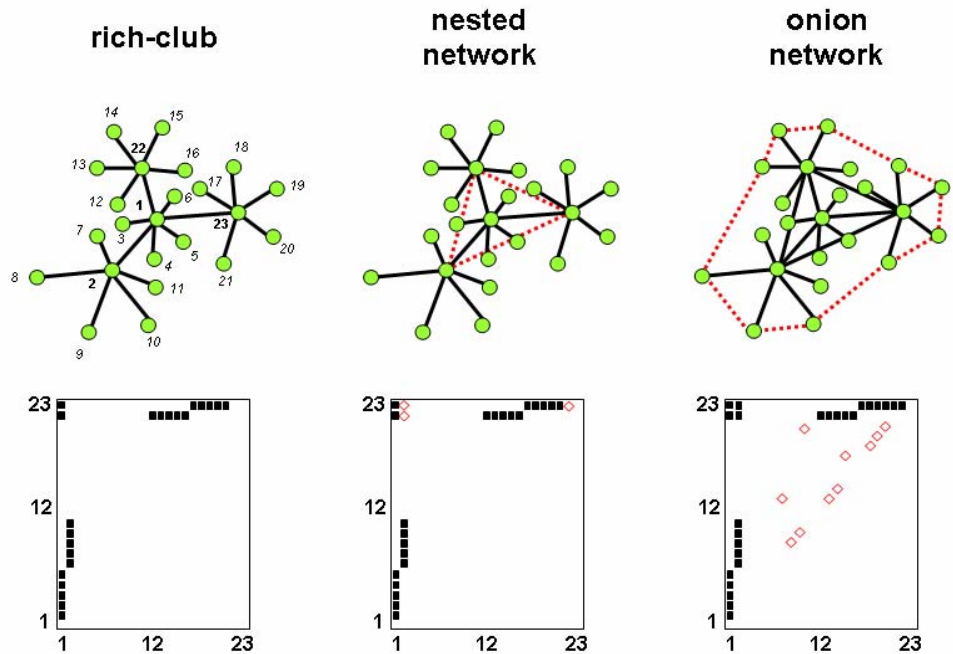

FIG. 2. Illustrations of various forms of core/periphery structures. The figure illustrates the differences between a network having a rich club (left side), having a nested structure (middle) and developing an onion-type topology (right side). Bottom figures give the adjacency matrices of the networks shown in the top row. Note that existing examples of nested networks were usually shown on bipartite networks. Networks of this illustrative figure are unipartite networks, since they show better the rich club and onion-type structures. Additionally, note that the connected hubs of the rich club became even more connected by adding the 3 dotted red edges on the middle panel (corresponding to the red open diamonds in the adjacency matrices), which provides a moderate increase in the nestedness of the network. Connection of the peripheral nodes by an additional 10 red edges on the right panel turns the nested network to an onion network having a core and an outer layer. Lastly, note that the rich club network already has a moderately nested structure, and both the nested network and the onion network have a rich club. Larger onion networks have multiple peripheral layers. Adapted from [131].